\documentclass[english]{article}
\usepackage{mathptmx}

\usepackage[T1]{fontenc}
\usepackage[utf8]{inputenc}
\usepackage{geometry}
\usepackage{amsmath}
\usepackage{amssymb}
\usepackage{babel}
\begin{document}

\title{Properties of Hubbard models with degenerate localised single-particle
eigenstates}

\author{Andreas Mielke\\
Institut für Theoretische Physik, Ruprecht Karls Universität\\
 Philosophenweg 19, D-69120 Heidelberg, F.R. Germany }

\date{\today}
\maketitle
\begin{abstract}
We consider the repulsive Hubbard model on a class of lattices or
graphs for which there is a large degeneracy of the single-particle
ground states and where the projector onto the space of single-particle
ground states is highly reducible. This means that one can find a
basis in the space of the single-particle ground states such that
the support of each single-particle ground state belongs to some small
cluster and these clusters do not overlap. We show how such lattices
can be constructed in arbitrary dimensions. We construct all multi-particle
ground states of these models for electron numbers not larger than
the number of localised single-particle eigenstates. We derive some
of the ground state properties, esp. the residual entropy, \emph{i.e.}
the finite entropy density at zero temperature. 
\end{abstract}

\section{Introduction}

The physics of strongly-correlated Fermi systems is one of the most
exciting branches of condensed matter theory. The most extreme case
of strong correlations occurs in systems where the single-particle
problem has a large degeneracy. Such systems with flat bands have
been studied intensively during the last 20 years. One of the more
prominent examples is the Hubbard model with a lowest flat band. The
first examples were discussed by Tasaki and by the present author
20 years ago \cite{Mielke1991,Tasaki92,Mielke1993}. The aim of the
authors was a rigorous, complete classification of the ground states.
Among other results, it was shown that under certain conditions the
Hubbard model has a unique ferromagnetic ground state. The final goal
was the proof of the existence of metallic ferromagnetism in these
and similar models.

If a translationally-invariant system has a flat band, not only the
Bloch states but also the Wannier states are eigenstates of the single-particle
Hamiltonian. This means that there are localised single-particle eigenstates.
Some of the physical properties of these systems can be related to
the existence of localised eigenstates. Most of the proofs used to
derive exact results for ferromagnetism in Hubbard models with flat
bands rely on the fact that such localised basis states exist. On
the other hand, the existence of metallic ferromagnetism, which was
always one of the goals, is not related to localised eigenstates.
One class of models where one could have metallic ferromagnetism is
constructed from the flat band models with additional hopping terms
to lift the degeneracy of the degenerate band \cite{Tasaki96,Tanaka2003}.
Another class of models have partially flat bands and thus the localised
Wannier states are not eigenstates \cite{Mielke1999a,Mielke1999}.

In the models mentioned so far the localised states, when they exist,
overlap. To be precise, the representation of the projector onto the
subspace of all degenerate single-particle ground states in position
space, the single-particle density matrix, is irreducible. This was
essential for the proofs in \cite{Mielke1993,Mielke1999,Mielke1999a}.
In \cite{Mielke1999} it was shown that if and only if the single-particle
density matrix is irreducible, the multi-particle ground state at
a special density, where each of the single-particle eigenstates is
filled with exactly one electron, is ferromagnetic and unique up to
the usual $(2S+1)$ fold spin degeneracy. 

Other models which recently caught some interest have strict\-ly
localised eigenstates in the sense that different localised single-particle
eigenstates do not overlap, see \emph{e.g.} Batista and Shastry \cite{Batista2003},
Dzerkho\emph{ et al.} \cite{Derzhko2009}, Maksymenko \emph{et al.}
\cite{Maksymenko2011}, some of the examples by Schmidt \emph{et al.}
\cite{Schmidt2006}, and the references therein. These authors consider
examples of decorated lattices in one or two dimensions. In these
systems, the physical properties are strongly influenced by the existence
of localised eigenstates. The system has a high degeneracy for the
multi-par\-tic\-le ground states. The entropy density at zero temperature
is finite. Since the single-particle eigenstates are localised the
system is likely to be non-metallic. In some cases the ground states
are Wigner crystals. The system is paramagnetic, not ferromagnetic.

The goal of the present paper is to provide a complete description
of the class of lattices with degenerate single-particle eigenstates
which fall into non-overlapping subsets. This is done using the projector
onto the subspace of degenerate single-particle ground states in position
space, the single-particle density matrix. The class of lattices with
single-particle eigenstates which fall into non-overlapping subsets
has a reducible single-particle density matrix, in contrast to the
models discussed in \cite{Mielke1999,Mielke1999a}. We discuss some
of the properties of the Hubbard model on such lattices and we give
a large class of examples of such lattices which can be constructed
explicitly. The examples in \cite{Batista2003,Derzhko2009,Maksymenko2011}
belong to this class as well. The construction is possible in arbitrary
dimensions. We construct explicitly all ground states of these models
for sufficiently low densities of states -- the flat band must be
at most half filled -- and we calculate the entropy density at zero
temperature. We prove that there is no long-range order in these models.

The paper is organised as follows: In Sect. \ref{sec:Classification-of-the}
we define the class of lattices we are looking at. They are defined
by some properties of the projector onto the single-particle ground
states of the system. We give a general description how examples of
such lattices can be constructed explicitely in arbitrary dimensions.
In the Sect. \ref{sec:Ground-state-properties}, we state and proof
our main result concerning the multi-particle ground states in such
models. Sect. \ref{sec:Summary-and-Outlook} contains a summary and
an outlook.

\section{Classification of the single-particle problem\label{sec:Classification-of-the}}

In this paper we consider a general fermionic Hubbard model 
\begin{equation}
H=H_{\textrm{hop}}+H_{\textrm{int}}
\end{equation}
 where 
\begin{equation}
H_{\textrm{hop}}=\sum_{\{x,y\}\in E,\sigma}t_{xy}c_{x\sigma}^{\dagger}c_{y\sigma}
\end{equation}
 and 
\begin{equation}
H_{\textrm{int}}=\sum_{x\in V}U_{x}n_{x\uparrow}n_{x\downarrow}
\end{equation}
 on a lattice or, more generally, on a connected graph $G=(V,E)$
with a set of vertices $V$ and edges $E$ connecting the vertices.
$t_{xy}$ are the hopping matrix elements, $U_{x}>0$ is the local
repulsive interaction. Two vertices $x$ and $y$ are connected by
an edge $e=\{x,y\}$ if and only if $t_{xy}\neq0$. In this section,
we consider first the single-particle problem in order to define the
class of models (or lattices) we are dealing with. 

We consider the case where $H_{{\rm hop}}$ has a highly degenerate
single-particle ground state with eigenenergy $\epsilon_{d}$. The
degeneracy is $N_{d}$. $G$ does not need to be translationally invariant.
In the case of a translationally-invariant lattice, we assume that
the system has at least one degenerate energy band at the bottom of
the spectrum. 

Let $B=\{\psi_{i}(x),\, i=1\ldots N_{d}\}$ be an arbitrary orthonormal
basis in the subspace of the degenerate lowest eigenstates of the
matrix $T=(t_{xy})_{x,y\in V}$. We assume that $t_{xy}$ are real,
however a generalisation to complex $t_{xy}$ is straight forward.
Complex $t_{xy}$ have also been discussed in the context of flat
bands, see \emph{e.g.} \cite{Gulacsi2008}. For real $t_{xy}$ we
choose the basis $B$ to be real as well. The single-particle density
matrix of these states is 
\begin{equation}
\rho_{xy}=\sum_{i=1}^{N_{d}}\psi_{i}(x)\psi_{i}(y).
\end{equation}
 $\rho=(\rho_{xy})_{x,y\in V}$ is the projector onto the space spanned
by the single-particle ground states in position space. We showed
that if $\rho=(\rho_{xy})_{x,y\in V}$ is irreducible, the Hamiltonian
has ferromagnetic multi-particle ground states and that for special
particle numbers $N_{e}=N_{d}$, the ferromagnetic ground state is
unique up to the degeneracy due to the $SU(2)$ spin symmetry \cite{Mielke1999}.

In this paper, we consider the case where the single-particle density
matrix $\rho$ is highly reducible. $\rho$ should have the following
properties:
\begin{enumerate}
\item $\rho$ is reducible. It can be decomposed into $N_{r}$ irreducible
blocks $\rho_{k}$, $k=1,\ldots,N_{r}$. $N_{r}$ should be an extensive
quantity, \emph{i.e.} $N_{r}\propto N_{d}\propto|V|$, so that in
the thermodynamic limit the density of degenerate single-particle
ground states and the density of irreducible blocks are both finite. 
\item Let $V_{k}$ be the support of $\rho_{k}$, \emph{i.e.} the set of
vertices for which at least one element of $\rho_{k}$ does not vanish.
$\rho_{k,xy}=0$ if $x\notin V_{k}$ or $y\notin V_{k}$. One has
$V_{k}\cap V_{k'}=\emptyset$ if $k\neq k'$ because of the fact that
$\rho_{k}$ are irreducible blocks of the reducible matrix $\rho$
and $\bigcup_{k}V_{k}\subseteq V$. 
\item We choose the $B$ such that the support of each basis states $\psi_{i}(x)$
is a subset of exactly one $V_{k}$. We denote the number of states
belonging to the cluster $V_{k}$ as $\nu_{k}$. One has $\sum_{k}\nu_{k}=N_{d}$. 
\item $\nu_{{\rm max}}=\max_{k}\{\nu_{k}\}$ is $O(1)$, \emph{i.e.} not
an extensive quantity. 
\end{enumerate}
If $G$ represents a translationally-invariant lattice, only one or
a few blocks belong to one elementary cell and the $\nu_{k}$ belong
to classes where, within one class, all $\nu_{k}$ are the same due
to translational invariance.

On each block, since $\rho_{k}$ is irreducible, the results obtained
in \cite{Mielke1999} apply.

\subsection{Lattices with such properties }

The lattices in \cite{Batista2003,Derzhko2009,Maksymenko2011} have
the properties mentioned above. We now give a more general construction
for a large class of lattices in arbitrary dimensions which have these
properties. Let us mention that these are only examples and that many
other lattices with reducible $\rho_{xy}$ exist.

Our starting point for the construction of a large class of lattices
or graphs with these properties is an arbitrary lattice or graph $\tilde{G}=(\tilde{V},\tilde{E})$.
$\tilde{V}$ is the set of vertices of $\tilde{G}$. $\tilde{E}$
is the set of edges of $\tilde{G}$. We consider only simple graphs,
\emph{i.e.} each edge is a set of exactly two vertices. If an edge
$\{x,y\}\in\tilde{E}$ exists, the two vertices $x$ and $y$ are
connected. For our construction, we decompose the vertex set $\tilde{V}$
into two disjoint subsets $\tilde{V}_{1}$ and $\tilde{V}_{2}$. As
a special case, $\tilde{V}_{2}$ may be empty. To each vertex $x\in\tilde{V}_{1}$
we associate a complete graph $K_{n}$ with $n$ vertices, $n\geq2$.
A complete graph is a graph where each vertex is connected with each
other vertex. $K_{2}$ is an edge, $K_{3}$ is a triangle, $K_{4}$
is a tetrahedron. These complete graphs form building blocks of the
new lattice. We denote these subgraphs as $K_{n}(x)$. For a discussion
of the Hubbard model on the complete graph the reader to referred
to \cite{Mielke1996} and the references therein.

We now construct the graph $G=(V,E)$ as follows: the vertex set $V$
is $V=V_{1}\bigcup V_{2}$ where $V_{2}=\tilde{V}_{2}$ and 
\begin{equation}
V_{1}=\bigcup_{x\in\tilde{V}_{1}}V(K_{n}(x))
\end{equation}
 where $V(K_{n}(x))$ is the vertex set of the complete graph $K_{n}(x)$
and $E(K_{n}(x))$ is its edge set. The edge set $E$ is constructed
as follows: 
\begin{equation}
E=E_{1}\cup E_{2}\cup E_{3}\cup E_{4}
\end{equation}
 
\begin{equation}
E_{1}=\bigcup_{x\in\tilde{V}_{1}}E(K_{n}(x))
\end{equation}
 
\begin{equation}
E_{2}=\{\{x,y\}\in\tilde{E}:\, x,y\in V_{2}\}
\end{equation}
 
\begin{equation}
E_{3}=\bigcup_{x\in\tilde{V}_{1,}y\in V_{2},\{x,y\}\in\tilde{E}}\{\{z,y\}:\, z\in V(K_{n}(x))\}
\end{equation}
 
\begin{align}
E_{4}=\bigcup_{x\in\tilde{V}_{1,}x'\in\tilde{V}_{1},\{x,x'\}\in\tilde{E}} & \{\{z,z'\}:\, z\in V(K_{n}(x)),\nonumber \\
 & z'\in V(K_{n}(x'))\}.\label{eq:E4}
\end{align}

In words: if $x,x'\in\tilde{V}_{1}$ are nearest neighbours in $\tilde{G}$,
we connect each vertex of $K_{n}(x)$ with each vertex of $K_{n}(x')$.
If $x\in\tilde{V}_{1}$, $y\in V_{2}$ are nearest neighbours in $\tilde{G}$,
we connect each vertex of $K_{n}(x)$ with $y$.

We associate with each edge in $E_{i}$ a hopping matrix element $t_{i}\geq0$
and denote this new hopping matrix by $T$. We add to $T$ on-site
energies $\bar{t}_{2}$ for the vertices in $V_{2}$. The eigenstates
and the spectrum of this single-particle hopping matrix can be constructed
from the eigenstates and the spectrum of the adjacency matrix $A(\tilde{G)}$
of the original graph $\tilde{G}$, $A(\tilde{G})=(a_{xy})_{x,y\in V(\tilde{G})}$
where $a_{xy}=1$ if $\{x,y\}\in\tilde{E}$, 0 otherwise. To show
this, we write 
\begin{equation}
A(\tilde{G})=\left(\begin{array}{cc}
A_{11} & A_{12}\\
A_{21} & A_{22}
\end{array}\right)
\end{equation}
 where $A_{ij}$ contains the matrix elements of $A(\tilde{G})$ connecting
vertices of $\tilde{V_{i}}$ to $\tilde{V}_{j}$. We introduce the
identity matrices $E_{i}=(\delta_{z,z'})_{z,z'\in V_{i}}$ on $V_{i}$
and the matrix $B=(b_{zx})_{z\in V_{1},x\in\tilde{V_{1}}}$, $b_{zx}=1$
if $z\in K_{n}(x)$ and $b_{zx}=0$ otherwise. $B$ maps the vertices
in $V_{1}$ to the vertices in $\tilde{V}_{1}$. Then, the new hopping
matrix $T$ can be written as
\begin{equation}
T=\left(\begin{array}{cc}
t_{1}(BB^{t}-E_{1})+t_{4}BA_{11}B^{t} & t_{3}BA_{12}\\
t_{3}A_{21}B^{t} & t_{2}A_{22}+\bar{t}_{2}E_{2}
\end{array}\right).
\end{equation}
Note that $B^{t}B=n\tilde{E}_{1}$, where $\tilde{E}_{1}$ is the
identity matrix on $\tilde{V}_{1}$. We now construct all eigenstates
of $T$. One class of eigenstates of $T$ are 
\begin{equation}
\psi=\left(\begin{array}{c}
\psi_{0}\\
0
\end{array}\right),\quad\psi_{0}\in\mbox{kernel}(B^{t})\label{eq:gsT}
\end{equation}
with the eigenvalue $-t_{1}$. A basis in this eigenspace is formed
by all eigenstates with eigenvalue -1 of $K_{n}(x)$ for all $x\in\tilde{V}_{1}$.
$K_{n}$ has one eigenvalue $n-1$ belonging to the eigenstate $\phi_{0}=(1,1,\ldots,1)^{t}$
and $n-1$ eigenvalues -1 belonging to eigenstates orthogonal to $\phi_{0}$,
these are elements of the kernel of $B^{t}$. The second class of
eigenstates of $T$ are of the form
\begin{equation}
\psi=\left(\begin{array}{c}
aB\psi_{1}\\
\psi_{2}
\end{array}\right)\;\text{where}\: A(\tilde{G})\left(\begin{array}{c}
\psi_{1}\\
\psi_{2}
\end{array}\right)=\tilde{a}\left(\begin{array}{c}
\psi_{1}\\
\psi_{2}
\end{array}\right).\label{eq:otherEigenstates}
\end{equation}
We obtain 
\begin{equation}
T\psi=\left(\begin{array}{c}
t_{1}(n-1)aB\psi_{1}+t_{4}anBA_{11}\psi_{1}+t_{3}BA_{12}\psi_{2}\\
t_{3}A_{21}an\psi_{1}+t_{2}A_{22}\psi_{2}+\bar{t}_{2}\psi_{2}
\end{array}\right)
\end{equation}
where we made use of $B^{t}B=nE_{1}$. We now let $a=n^{-1/2}$, $t_{3}=t_{4}/a$,
$t_{2}=t_{3}/a$, and $\bar{t}_{2}=t_{1}(n-1)$. Then, we obtain $T\psi=[t_{1}(n-1)+t_{4}\tilde{a}]\psi$.
We choose $t_{1}$ and $t_{4}$ so that $t_{1}n+t_{4}\tilde{a}_{{\rm min}}>0$
where $\tilde{a}_{{\rm min}}$ is the lowest eigenvalue of $A(\tilde{G})$.
Since $\tilde{a}_{{\rm min}}\geq-d_{{\rm max}}(\tilde{G})$, which
is the maximal degree of $\tilde{G},$ it is sufficient to choose
$t_{4}<t_{1}n/d_{{\rm max}}(\tilde{G})$. Then the ground state of
$T$ has a lowest eigenvalue $-t_{1}$ with degeneracy $N_{d}=(n-1)|\tilde{V}_{1}|$
and the ground states are given by (\ref{eq:gsT}). By construction,
there are $|\tilde{V}$| eigenstates with eigenvalues above $-t_{1}$
and since $N_{d}+|\tilde{V}|=|V|$ the construction yields all eigenstates
of $T$. 

This construction is valid even if $V_{2}$ is empty. 

Let us remark that if $\tilde{G}$ is a translationally-invariant
lattice with $r$ energy bands and $G$ is translationally invariant
as well, then $G$ has $r+n-1$ energy bands and the $n-1$ additional
energy bands are flat and degenerate with eigenenergy $-t_{1}$.

This construction yields a system in which the matrix $\rho$ decomposes
into $|\tilde{V}_{1}|$ blocks, one for each $x\in\tilde{V}_{1}$.
The $\nu_{x}$ are all the same and their value is $n-1$.

The construction can be generalized, we may choose different values
of $n$ for different $x\in\tilde{V}$.

\section{Ground state properties\label{sec:Ground-state-properties}}

We are now ready to state our main results for the Hubbard model with
a lowest single-particle eigenenergy $\epsilon_{d}$ which is $N_{d}$-fold
degenerate and for which the projector onto the eigenspace of $\epsilon_{d}$
fulfills the properties of Sect. \ref{sec:Classification-of-the}.

\paragraph*{Theorem. }

For Hubbard models with a lowest single-particle eigenenergy $\epsilon_{d}$
which is $N_{d}$-fold degenerate and for which the projector onto
the eigenspace of $\epsilon_{d}$ fulfills the properties of Sect.
\ref{sec:Classification-of-the}, the following results hold for $N_{e}\leq N_{d}$:
\begin{enumerate}
\item The ground state energy is $\epsilon_{d}N_{e}$.
\item Let $A_{x}$ be an arbitrary local operator, \emph{i.e.} an arbitrary
combination of the four creation and annihilation operators $c_{x\sigma}^{\dagger}$
and $c_{x\sigma}$. The correlation function $\rho_{A,xy}=\langle A_{x}A_{y}\rangle-\langle A_{x}\rangle\langle A_{y}\rangle$
has a finite support for any fixed $x$ and vanishes if $x$ and $y$
are out of different clusters $V_{k}$. The system has no long-range
order.
\item The system is paramagnetic.
\item The entropy at zero temperature $S(c)$ is an extensive quantity,
$S(c)=O(N_{e})$. It increases as a function of $c=N_{e}/N_{d}$ from
$0$ for $c=0$ to some maximal value $S_{{\rm max}}\geq\sum_{k}[(\nu_{k}-1)\ln2+\ln(\nu_{k}+2)]$
and then decays to $S(1)=\sum_{k}\ln(\nu_{k}+1)$.
\end{enumerate}

\paragraph*{Proof. }

We first construct a suitable basis. The different clusters $V_{k}$
are completely decoupled as long as ground states are considered.
We therefore discuss first a single cluster. According to \cite{Mielke1999},
it is possible to choose a single-particle basis $B_{k}$ (not orthonormal)
for the space of degenerate single-particle ground states on the cluster
$V_{k}$ with the following properties:
\begin{enumerate}
\item For each basis state $\psi_{i}$ there exists a vertex set $V_{k(i)}$
so that the support of $\psi_{i}$ is a subset of $V_{k(i)}$. 
\item For each basis state $\psi_{i}$ there exists a unique $x_{i}\in V_{k(i)}$
such that $\psi_{i}(x_{i})>0$ and $\psi_{i}(x_{j})=0$ for all $i\neq j$,
$j=1,\ldots,|B_{k}|$. The set $V_{k(i)}\setminus\{x_{i},\, i=1,\ldots,|B_{k}|\}$
is not empty. 
\end{enumerate}
Since the set $V_{k(i)}\setminus\{x_{i},\, i=1,\ldots,|B_{k}|\}$
is not empty, these states overlap and are not orthogonal. Let $\nu_{k}=|B_{k}|$.
Using such a basis $B_{k}$, we can put electrons into the different
basis states with the condition that one single-particle basis state
contains at most one electron. In a state with two electrons in a
basis state $\psi_{i}$ we have a double occupancy on $x_{i}$. Since
all other basis states vanish on $x_{i}$, it is not possible to get
rid of that double occupancy due to some linear combinations of these
states. Therefore, any state with a doubly-occupied $\psi_{i}$ has
a non-vanishing interaction energy and cannot be a ground state. Note
that the absence of doubly occupied $\psi_{i}$ is a necessary but
not a sufficient condition for a ground state, since it does not exclude
double occupancies on the lattices sites in $V_{k(i)}\setminus\{x_{i},\, i=1,\ldots,|B_{k}|\}$.
This construction was used in \cite{Mielke1999} to show that if there
are $\nu_{k}$ electrons on the cluster, they all have the same spin
$S_{k}=\nu_{k}/2$ and that the degeneracy of the ground state on
the cluster is $2S_{k}+1=\nu_{k}+1$. The trivial case is one electron
on the cluster, where the degeneracy is $2\nu_{k}$. For electron
numbers $n_{k}$ with $1<n_{k}<\nu_{k}$ it may be difficult to calculate
the ground state degeneracies. But for $\nu_{k}\leq2$ we have a complete
description of all ground states in the cluster $V_{k}$. 

It is trivial to generalise this argument to the entire lattice using
the basis $\bigcup_{k}B_{k}$. The states in the different $B_{k}$
can be filled independently. The lattices studied in \cite{Batista2003,Derzhko2009,Maksymenko2011}
all belong to the class with $\nu_{k}\leq2$. In \cite{Maksymenko2011},
it was assumed that the construction above yields all of the ground
states and numerical results were presented to confirm that. Our argument
is a rigorous proof of this statement.

We now come to the four statements in the theorem. The first point
is trivial. The ground state energy is $\epsilon_{d}N_{e}$, since
states with that energy minimise both the kinetic energy and the interaction.
The ground states have no doubly-occupied sites.

For the proof of the next statements, we use a grand-canonical formulation.
Let
\begin{equation}
Z(z,\{a_{x},\, x\in V\})=\langle z^{N_{e}}\exp(\sum_{x}a_{x}A_{x})\rangle
\end{equation}
be the generating function for correlation functions containing the
operators $A_{x}$. $N_{e}$ is the number operator. $\langle\cdot\rangle$
denotes the ground state expectation value for arbitrary electron
numbers $\leq N_{d}$. Since the system decomposes into clusters,
the generating function can be written as 
\begin{equation}
Z(z,\{a_{x},\, x\in V\})=\prod_{k}Z(z,\{a_{x},\, x\in V_{k}\},\rho_{k})
\end{equation}
where 
\begin{equation}
Z(z,\{a_{x},\, x\in V_{k}\},\rho_{k})=\langle z^{N_{e}}\exp(\sum_{x}a_{x}A_{x})\rangle_{k}
\end{equation}

$\langle\cdot\rangle_{k}$ denotes the ground state expectation value
on the cluster $V_{k}$. We have
\begin{equation}
\langle A_{x}\rangle=\frac{\partial}{\partial a_{x}}\left.\ln Z(z,\{a_{x},\, x\in V\})\right|_{a_{x}=0\,\forall x}
\end{equation}
\begin{equation}
\langle A_{x}A_{y}\rangle=\frac{\partial^{2}}{\partial a_{x}\partial a_{y}}\left.\ln Z(z,\{a_{x},\, x\in V\})\right|_{a_{x}=0\,\forall x}
\end{equation}
Since $\ln Z(z,\{a_{x},\, x\in V\})=\sum_{k}\ln Z(z,\{a_{x},\, x\in V_{k}\},\rho_{k})$
one has $\langle A_{x}A_{y}\rangle=\langle A_{x}\rangle\langle A_{y}\rangle$
if $x$ and $y$ are out of different clusters $V_{k}$. Thus, $\rho_{A,xy}$
vanishes if $x$ and $y$ are out of different clusters. Since this
statement holds for any $z$, it holds as well for a fixed particle
number $N_{e}$. 

The third point follows from the second if we take for $A_{x}$ the
local spin-operators. To be more explicit, let us calculate the expectation
value of the total spin. It can be written as $\langle\vec{S}^{2}\rangle=\sum_{k}\langle\vec{S}_{k}^{2}\rangle$
where $\vec{S}_{k}=\sum_{x\in V_{k}}\vec{S}_{x}$ is the spin operator
on the cluster $k$. A trivial upper bound for $\langle\vec{S}_{k}^{2}\rangle$
is $\frac{1}{4}\nu_{k}(\nu_{k}+2)$. A trivial lower bound for $\langle\vec{S}_{k}^{2}\rangle$
is $\frac{3}{4}N_{k}$, where $N_{k}$ is the number of electrons
on the cluster $V_{k}$. Therefore, we obtain $\frac{3}{4}N_{e}\leq\langle\vec{S}^{2}\rangle=S_{{\rm tot}}(S_{{\rm tot}}+1)\leq\frac{1}{4}\sum_{k}\nu_{k}(\nu_{k}+2)\leq\frac{1}{4}\nu_{{\rm max}}(\nu_{{\rm max}}+2)N_{r}$.
Therefore, $S_{{\rm tot}}$ is not an extensive quantity, the system
is not ferromagnetic. The maximum value of $S_{{\rm tot}}$ occurs
for $N_{e}=N_{d}$, and is given by $S_{{\rm tot}}(S_{{\rm tot}}+1)=\frac{1}{4}\sum_{k}\nu_{k}(\nu_{k}+2)$.
This proves the third point in the theorem. 

We come now to the fourth point of the theorem. We will calculate
the entropy density by calculating the ground state degeneracy of
the system. The fact that this can be done for lattice models with
finite range interactions has been proven by Aizenman and Lieb \cite{Aizenman1981}.
They pointed out that there is a problem when interchanging the limit
$T\rightarrow0$ and the thermodynamic limit. In order to calculate
the entropy density at zero temperature, one has to calculate the
ground state degeneracy for all possible boundary conditions. This
is possible in our case since the lattice decomposes into finite clusters,
so that the boundary has no effect on the result.

To calculate the entropy at zero temperature, let us now calculate
the grand canonical partition function. Since the problem decomposes
into a set of clusters $V_{k}$, the contribution of these multi-particle
ground states to the grand canonical partition function is a product
of the partition functions of these clusters. 
\begin{equation}
Z(z)=\prod_{k}Z(z,\rho_{k}).
\end{equation}

The general form of $Z(z,\rho_{k})$ is 
\begin{equation}
Z(z,\rho_{k})=\sum_{j=0}^{\nu_{k}}p_{j}^{\nu_{k}}z^{j}
\end{equation}
 where $p_{j}^{\nu}$ is the number of states with $j$ electrons
on a cluster with $\nu$ states. One has $p_{0}^{\nu}=1$, $p_{1}^{\nu}=2\nu$,
$p_{\nu}^{\nu}=\nu+1$, $p_{j}^{\nu}\geq(j+1)\left(\begin{array}{c}
\nu\\
j
\end{array}\right)$ for $1<j<\nu$. The lower limit for $p_{j}^{\nu}$ is the number
of fully-polarised states with $j$ electrons on a cluster with $\nu$
states.

From $Z$ we obtain the grand canonical potential $\Omega$. The general
from of $\Omega$ is 
\begin{equation}
\Omega=-\beta^{-1}\sum_{k}\ln Z(z,\rho_{k}).
\end{equation}
 The entropy is

\begin{equation}
S(z)=-\frac{\partial\Omega}{\partial T}=\sum_{k}\ln Z(z,\rho_{k})-z\ln z\sum_{k}\frac{d}{dz}\ln Z(z,\rho_{k}).
\end{equation}

Let us introduce $c(z)=N_{e}(z)/N_{r}$. Since $N_{e}(z)=\frac{\partial\Omega}{\partial\mu}$
we obtain $c(z)=\frac{1}{N_{r}}\sum_{k}N_{k}(z)$ where 
\begin{equation}
N_{k}(z)=\frac{z}{Z(z,\rho_{k})}\frac{\partial Z(z.\rho_{k})}{\partial z}
\end{equation}
 is the number of particles on the cluster $V_{k}$. One has $S(z)=\sum_{k}S_{k}(z)$,
$S_{k}(z)=\ln Z(z,\rho_{k})-N_{k}(z)\ln z$.

$c(z)$ is a strongly monotonically increasing function of $z$. For
the derivative of the entropy, we obtain 
\begin{eqnarray}
\frac{dS}{dz} & = & -\ln z\left[\sum_{k}\frac{d\ln Z(z,\rho_{k})}{dz}+z\sum_{k}\frac{d^{2}\ln Z(z,\rho_{k})}{dz^{2}}\right]\nonumber \\
 & = & -\ln z\left[1+z\frac{d}{dz}\right]\frac{1}{z}\sum_{k}N_{k}(z)\\
 & = & -N_{r}\ln z\frac{dc}{dz}.\nonumber 
\end{eqnarray}
 Since $\frac{dc}{dz}>0$, the only maximum of $S(z)$ occurs m at
$z=1$. The value is $S(z=1)=\sum_{k}\ln Z(1,\rho_{k})$. For small
values of $z$ we have

\begin{equation}
Z(z,\rho_{k})=1+2\nu_{k}z+O(z^{2})
\end{equation}
 and therefore

\begin{equation}
S(z)=2z(1-\ln z)N_{d}(1+O(z))
\end{equation}

In the limit $z\rightarrow\infty$, at the maximal density, the degeneracy
in the cluster $\rho_{k}$ is $\nu_{k}+1$. The total degeneracy is
$\prod_{k}(\nu_{k}+1)$ and the entropy is $S(z\rightarrow\infty)=\sum_{k}\ln(\nu_{k}+1)$.
Therefore, the entropy increases monotonically to it's maximum at
$z=1,$ $S(z=1)=\sum_{k}\ln Z(1,\rho_{k})$ and then decays monotonically
to $S(z\rightarrow\infty)=\sum_{k}\ln(\nu_{k}+1)$. Using $p_{j}^{\nu}\geq(j+1)\left(\begin{array}{c}
\nu\\
j
\end{array}\right)$ we obtain the lower bound for $S(z=1)$ in point 4 of the theorem.
Since $c(z)$ is strongly monotonically increasing, these properties
hold for $S(c)$ as well.

\subsection{Examples}

For clusters with $\nu_{k}=1$ one has 
\begin{equation}
Z(z,\rho_{k})=1+2z.\label{eq:nuk=00003D00003D1}
\end{equation}
 For $\nu_{k}=2$ one obtains 
\begin{equation}
Z(z,\rho_{k})=1+4z+3z^{2}\label{eq:nuk=00003D00003D2}
\end{equation}
 since there is one state on the cluster $V_{k}$ with no particles,
four states with one particle and three states with two particles.
For $\nu_{k}=3$ one has $2\nu_{k}=6$ states with one electron and
$\nu_{k}+1=4$ states with three electrons. Two electrons on the cluster
can form a triplet state or a singlet state. One gets $3\nu_{k}=9$
triplet states and between 0 and 2 singlet states. In the basis $B_{k}$
none of the basis states can be doubly occupied. Therefore one has
only three different pairs which could form a singlet. But, since
$\rho_{k}$ is irreducible, at most two different pairs without a
doubly occupied site can be constructed. Therefore, for $\nu_{k}=3$
we obtain 
\begin{equation}
Z(z,\rho_{k})=1+6z+(9+s_{k})z^{2}+4z^{3}\label{eq:nuk=00003D00003D3}
\end{equation}

where $s_{k}$ is the number of possible singlets on the cluster $\rho_{k}$,
which can be 0, 1, or 2.

If $\nu_{k}\leq3$ for all $k$, the total partition function is thus
\begin{eqnarray}
Z(z) & = & (1+2z)^{N_{1,0}}(1+4z+3z^{2})^{N_{2,0}}\nonumber \\
 &  & (1+6z+9z^{2}+4z^{3})^{N_{3,0}}\nonumber \\
 &  & (1+6z+10z+4z^{3})^{N_{3,1}}\label{eq:Z123}\\
 &  & (1+6z+11z^{2}+4z^{3})^{N_{3,2}}\nonumber 
\end{eqnarray}
 where $N_{\nu,s}$ is the number of clusters with $\nu_{k}=\nu$
and $s$ possible singlet states. From (\ref{eq:Z123}) one obtains
the grand canonical potential 
\begin{eqnarray}
\frac{\Omega}{N_{r}} & = & -\beta^{-1}(n_{1,0}\ln(1+2z)+n_{2,0}\ln(1+4z+3z^{2})\nonumber \\
 &  & +n_{3,0}\ln(1+6z+9z^{2}+4z^{3})\nonumber \\
 &  & +n_{3,1}\ln(1+6z+10z^{2}+4z^{3})\label{eq:Omega123}\\
 &  & +n_{3,2}\ln(1+6z+11z^{2}+4z^{3}))\nonumber 
\end{eqnarray}
 where $n_{\nu,s}=N_{\nu,s}/N_{r}$, $n_{1,0}+n_{2,0}+n_{3,0}+n_{3,1}+n_{3,2}=1$.

Let us mention that all of the one-dimensional lattices treated in
\cite{Maksymenko2011} belong to this class with $n_{2,0}=1$, \emph{i.e.}
$N_{1,0}=0$ and $N_{3,s}=0$ for $s=0,1,2$. 

Inverting $c(z)$ we obtain $z$ as a function of $c$ and then $S(c)$.
There are two cases where $c(z)$ can be inverted easily, the case
where all $\nu_{k}=1$ and the case where all $\nu_{k}=2$. In the
first case one obtains a linear equation for $z(c)$, in the second
case a quadratic. The one-dimensional models in \cite{Maksymenko2011}
are a special case for $\nu_{k}=2$. The result given there is valid
for any model where all clusters have $\nu_{k}=2$. For the case $\nu_{k}\leq3$,
(\ref{eq:Omega123}), one obtains 
\begin{eqnarray}
c & = & n_{1,0}\frac{2z}{1+2z}+n_{2,0}\frac{4z+6z^{2}}{1+4z+3z^{2}}\nonumber \\
 &  & +n_{3,0}\frac{6z+18z^{2}+12z^{3}}{1+6z+9z^{2}+4z^{3}}\nonumber \\
 &  & +n_{3,1}\frac{6z+20z^{2}+12z^{3}}{1+6z+10z^{2}+4z^{3}}\label{eq:c123}\\
 &  & +n_{3,2}\frac{6z+22z^{2}+12z^{3}}{1+6z+11z^{2}+4z^{3}}.\nonumber 
\end{eqnarray}

One has $c\leq n_{1}+2n_{2}+3(n_{3,0}+n_{3,1}+n_{3,2})=1+n_{2}+2(n_{3,0}+n_{3,1}+n_{3,2})$,
where the upper limit is reached for $z\rightarrow\infty$. In this
limit, the entropy takes the value 
\begin{equation}
S=N_{r}(n_{1,0}+2n_{2,0}+3(n_{3,0}+n_{3,1}+n_{3,2})).
\end{equation}
The maximum 
\begin{align}
S=N_{r} & (n_{1,0}\ln3+n_{2,0}\ln8+n_{3,0}\ln20\nonumber \\
 & +n_{3,1}\ln21+n_{3,2}\ln22)\label{eq:Sbsp}
\end{align}
 of the entropy occurs at 
\begin{equation}
c=\frac{2}{3}n_{1,0}+\frac{5}{4}n_{2,0}+\frac{9}{5}n_{3,0}+\frac{38}{21}n_{3,1}+\frac{5}{4}n_{3,2}.
\end{equation}
For the examples constructed in Sect. \ref{sec:Classification-of-the},
the construction of this basis is easy. We simply use the states with
the properties $\psi_{i}(i)=1$, $\psi_{i}(n)=-1$, $\psi_{i}(j)=0$
for $j\neq i,n$, $i=1,\ldots,n-1$. The partition function is 
\begin{equation}
Z(z)=\prod_{x\in\tilde{V}_{1}}Z(z,K_{n}(x))=\prod_{n}Z(z,K_{n})^{N_{n}}.
\end{equation}
 It is sufficient to consider a single $K_{n}$. Let $V(K_{n})=\{1,\ldots,n\}$.
The cases $n\leq3$ correspond to the cases $\nu_{k}\leq2$ already
discussed above. $K_{4}$ has $\nu_{k}=3$. For $K_{4}$ it is possible
to construct two pairs of non-over\-lapp\-ing single-particle ground
states. Using the basis introduced above the two pairs are $\{\psi_{1},\psi_{2}-\psi_{3}\}$
and $\{\psi_{1}-\psi_{2},\psi_{3}\}$. Therefore, we have $N_{3,0}=N_{3,1}=0$
in that case, and $Z(z,K_{4})=1+6z+11z^{2}+4z^{3}$. For larger values
of $n$ the number of non-trivial cases increases rapidly. In principle
it is possible to completely describe the multi-particle states as
well. We do not discuss these cases here.

The most important point in the discussion of these systems and of
the multi-particle ground states is that although this construction
allows for many different examples of solvable systems in arbitrary
dimensions (since in the above construction $\tilde{G}$ may be an
arbitrary lattice in arbitrary dimensions), the ground state properties
for $T=0$ and $N_{e}\leq N_{d}$ and the contribution of the ground
states to the low temperature properties of the system are that of
a collection of zero-dimensional systems. The properties do not depend
on the dimension of the lattice, but only on the number of different
subgraphs of type $K_{n}$ the lattice contains.

\section{Summary and Outlook\label{sec:Summary-and-Outlook}}

This paper yields a complete classification of all Hubbard models
for which the degeneracy $N_{d}$ of the single-particle ground states
is some finite fraction of the number of lattice sites and for which
the projector onto this subspace is highly reducible, \emph{i.e.}
where the number of irreducible submatrices $N_{r}$ of this projector
is some finite fraction of the number of lattice sites. Each subspace
lives on a local cluster and different clusters do not overlap. We
show how lattices with these properties can be constructed in arbitrary
dimensions and we derive some properties of the ground states of such
models for electron numbers $N_{e}\leq N_{d}$. Examples of such lattices
in one and two dimensions were previously presented by Batista and
Shastry \cite{Batista2003}, Maksymenko \emph{et al.} \cite{Maksymenko2011},
and others, see also the references therein. Maksymenko \emph{et al.}
\cite{Maksymenko2011} gave a rather complete discussion of the Hubbard
model on some one-dimensional lattices of this type.

The important point is that the ground states properties for such
models do not depend on the details of the lattice or on its dimensionality,
but only on the properties of the local clusters. Global properties
like the entropy at $T=0$ can be calculated. The behaviour of the
entropy is similar for all of these lattices. The entropy density
as a function of the density of particles grows from 0 to some maximum
and then decays to some finite value at $N_{e}=N_{d}$. Thus, the
one-dimensional lattices in \cite{Maksymenko2011} are ideal prototypes
of all these models, and no essentially new physics occurs in the
higher-dimensional models. Maksymenko \emph{et al.} \cite{Maksymenko2011}
discussed the case where the degeneracy within the cluster is lifted.
In that case the system still has a large ground state degeneracy,
finite entropy density, etc. In that case, the ground states still
are located on the small local clusters and the dimensionality of
the lattice remains unimportant. This may of course change if one
lifts the degeneracy by some small perturbation so that the lowest
bands are no longer strictly flat. For the discussion of the stability
of ferromagnetism the situation then becomes much more difficult,
see \cite{Tasaki96,Tanaka2003}. We expect that with such perturbations,
the dimensionality of the lattice becomes important as well and that
the analysis will be much more difficult. Nevertheless, in that case
new and interesting physics may occur.

Maksymenko \emph{et al.} \cite{Maksymenko2011} discussed not only
the Hubbard model but also the Heisenberg model on such lattices.
We expect that their results for the two one-dimensional models can
easily be generalised to the class of lattices described here as well.
We expect that, as for the Hubbard model, one obtains no new interesting
physics.

Another class of models which are closely related to spin systems
with antiferromagnetic exchange interactions are bosonic Hubbard models
with flat bands on similar lattices, see \emph{e.g.} \cite{Schmidt2006,Huber2010}
and the references therein. Bosonic Hubbard models can also be discussed
on the lattices presented here with similar results. 

In this paper, we only discussed the $T=0$ properties of these models.
For $T>0$, the situation becomes more complicated. The detailed structure
of the lattice and the dimensionality become important, since the
other single-particle eigenstates (\ref{eq:otherEigenstates}) depend
on the detailed lattice properties. 

\bibliographystyle{epj}

\end{document}